\documentstyle[11pt,fleqn]{article}
\def\soc{{\rm C}_{\rm 60}}
\def\rug{{\rm C}_{\rm 70}}

\def\la{\langle}
\def\ra{\rangle}
\def\beeq{\begin{equation}}
\def\eneq{\end{equation}}
\def\beeqa{\begin{eqnarray}}
\def\eneqa{\end{eqnarray}}

\addtocounter{section}{-1}
\begin{document}
\begin{center}

{\large {\bf {
Photophysical Properties in C$_{\bf 60}$ and Higher Fullerenes\\
} } }

Kikuo Harigaya\footnote[1]{E-mail address: 
\verb+harigaya@etl.go.jp+; URL: 
\verb+http://www.etl.go.jp/+\~{}\verb+harigaya/+}

{\sl Electrotechnical Laboratory, Tsukuba 305-8568, Japan}

\end{center}

\vspace{1cm}

\noindent
{\large {\bf Abstract}}\\
The optical excitations in C$_{60}$ and higher fullerenes, 
including isomers of C$_{76}$, C$_{78}$, and C$_{84}$, 
are theoretically investigated.  We use a tight binding 
model with long-range Coulomb interactions, treated by 
the Hartree-Fock and configuration-interaction methods.  
We find that the optical excitations in the energy region 
smaller than about 4 eV have most of their amplitudes at 
the pentagons.  The oscillator strengths of projected 
absorption almost accord with those of the total absorption.  
Next, off-resonant third order susceptibilities are investigated.
We find that third order susceptibilities of higher 
fullerenes are a few times larger than those of C$_{60}$.  
The magnitude of nonlinearity increases as the optical gap 
decreases in higher fullerenes.  The nonlinearity is nearly 
proportional to the fourth power of the carbon number when 
the onsite Coulomb repulsion is $2t$ or $4t$, $t$ being the 
nearest neighbor hopping integral.  This result, indicating 
important roles of Coulomb interactions, agrees with quantum 
chemical calculations of higher fullerenes.

\vspace{1cm}

\noindent
{\large {\bf 1. Phason Lines and Linear Absorption}}\\
Recently, the fullerene family C$_N$ with hollow cage structures 
has been intensively investigated.  A lot of optical experiments 
have been performed, and excitation properties due to $\pi$-electrons 
delocalized on molecular surfaces have been measured.  For example, 
the optical absorption spectra of $\soc$ and $\rug$ [1,2] have been 
reported, and the large optical nonlinearity of $\soc$ [3,4] has 
been found.  The absorption spectra of higher fullerenes (C$_{76}$, 
C$_{78}$, C$_{84}$, etc.) have also been obtained [5,6].  For 
theoretical studies, we have applied a tight binding model [7] to 
$\soc$, and have analyzed the nonlinear optical properties.  Coulomb 
interaction effects on the absorption spectra and the optical 
nonlinearity have been also studied [8].  We have found that the 
linear absorption spectra of $\soc$ and $\rug$ are well explained 
by the Frenkel exciton picture [9] except for the charge transfer 
exciton feature around the excitation energy 2.8 eV of the $\soc$ 
solids [2].  Coulomb interaction effects reduce the magnitude of 
the optical nonlinearity from that of the free electron calculation 
[8], and thus the intermolecular interaction effects have turned 
out to be important.

In the previous paper [10], we have extended the calculation of 
$\soc$ [9] to one of the higher fullerenes C$_{76}$.  We have 
discussed variations of the optical spectral shape in relation 
to the symmetry reduction from $\soc$ and $\rug$ to C$_{76}$: 
the optical gap decreases and the spectra exhibit a larger number 
of small structures in the dependences on the excitation energy.  
These properties seem to be natural when we take into account of 
the complex surface patterns composed of pentagons and hexagons.  
In order to understand the patterns clearly, the idea of the 
phason lines (Fig. 1) has been introduced [11] using the projection 
method on the honeycomb lattice plane [12].  There are twelve 
pentagons in C$_{76}$.  Six of them cluster on the honeycomb 
lattice, with one hexagon between the neighboring two pentagons.  
There are two groups of the clustered pentagons.  The phason line 
runs as if it divides the two groups.

We use the following Hamiltonian:
\beeq
H = H_0 + H_{\rm int}.
\eneq
The first term of eq. (1) is the tight binding model:
\beeq
H_0 = - t \sum_{\langle i,j \rangle, \sigma} 
(c_{i,\sigma}^\dagger c_{j,\sigma} + {\rm h.c.}),
\eneq
where $t$ is the hopping integral and $c_{i,\sigma}$ is an 
annihilation operator of a $\pi$-electron with spin $\sigma$
at the $i$th carbon atom of the fullerene.  It is assumed 
that $t$ does not depend on the bond length, because main 
contributions come from excitonic effects due to the strong 
Coulomb potential.  The results do not change so largely 
if we consider changes of hopping integrals by bond distortions.  
We assume the following form of Coulomb interactions among 
$\pi$-electrons:
\beeqa
H_{\rm int} &=& U \sum_i 
(c_{i,\uparrow}^\dagger c_{i,\uparrow} - \frac{1}{2})
(c_{i,\downarrow}^\dagger c_{i,\downarrow} - \frac{1}{2}) \nonumber \\
&+& \sum_{i \neq j} W(r_{i,j}) 
(\sum_\sigma c_{i,\sigma}^\dagger c_{i,\sigma} - 1)
(\sum_\tau c_{j,\tau}^\dagger c_{j,\tau} - 1),
\eneqa
where $r_{i,j}$ is the distance between the $i$th and $j$th atoms and
\beeq
W(r) = \frac{1}{\sqrt{(1/U)^2 + (r/r_0 V)^2}}
\eneq
is the Ohno potential.  The quantity $U$ is the strength of 
the on-site interaction, $V$ means the strength of the long-range
Coulomb interaction, and $r_0$ is the average bond length.

The model is treated by the Hartree-Fock approximation and the 
single excitation configuration interaction method, as was used 
in the previous papers [9,10].   In ref. 9, we have varied the 
parameters of the Coulomb interactions, and have searched for 
the data which reproduce overall features of experiments of 
$\soc$ and $\rug$ in solutions.  We have found that the common 
parameters, $U = 4t$ and $V = 2t$, are reasonable.  Thus, we use 
the same parameter set for higher fullerenes.  The quantity $t$ 
is about 2 eV as shown in ref. 9.  The Coulomb interaction strengths 
depend on the carbon positions.  We use the lattice coordinates 
which are obtained by the public program FULLER [13,14].  The 
optical spectra become anisotropic with respect to the orientation 
of the molecule against the electric field of light, as reported 
in the free electron model (H\"{u}ckel theory) [15].  We obtain 
numerical optical absorption spectra by summing the data of three 
cases, where the electric field of light is along the $x$-, $y$-, 
and $z$-axes.

We use a projection operator in order to extract contributions
to the optical spectra from a certain part of fullerenes.  
If we write the a projection operator to a part of lattice site
set as $P$, the oscillator strength between the ground state 
$|g\rangle$ and the excited state $|\kappa \rangle$ is written:
\beeqa
f_{\kappa,x} &=& E_\kappa [ | \la \kappa | P x P | g \ra |^2 \nonumber \\
&+& | \la \kappa | (1-P) x (1-P) | g \ra |^2 \nonumber \\
&+& \la g | P x P | \kappa \ra \la \kappa | (1-P) x (1-P) | g \ra \nonumber \\
&+& \la g | (1-P) x (1-P) | \kappa \ra \la \kappa | P x P | g \ra ],
\eneqa
where $E_\kappa$ is the excitation energy, and the electric
field is parallel with the $x$-axis.  In eq. (5), the first term
is the contribution from the projected part, and the three other
terms are the remaining part.  The total optical absorption
is calculated by the formula:
\beeq
\sum_\kappa \rho(\omega - E_\kappa) 
(f_{\kappa,x} + f_{\kappa,y} + f_{\kappa,z}),
\eneq
where $\rho(\omega) = \gamma/[\pi(\omega^2+\gamma^2)]$ 
is the Lorentzian distribution of the width $\gamma$.  
The projected absorption is calculated by eqs. (5) and (6).  
The projected part does not satisfy a sum rule.  So, this 
results in a singularity where excitation energy is large.  
We will discuss the optical spectra in the energy region
far from the singularity.

Figure 2 shows the molecular structures and optical spectra
of the $\rug$ molecule and C$_{76}$ with the $D_2$ symmetry, 
which have been found in experiments.  The black atoms are 
the carbons along the phason lines.  The hatched circles are 
the pentagonal carbons.  In C$_{70}$, the phason line runs 
along the ten carbons which are arrayed like a belt around 
the molecule.  In C$_{76}$, the phason line is located almost 
along the outer edge of the molecule (Fig. 1).  The total 
optical absorption is shown by the bold line, and the 
absorption from all the pentagonal carbons is shown by the 
thin line.  We find that the optical excitations in the energy 
region lower than 2$t$ are almost composed of the excitations 
at the pentagonal sites.  This property is common to $\rug$
and $D_2$-C$_{76}$, and also to the $T_d$-C$_{76}$ for which
the calculated data are not shown.  In higher energy regions,
the thin lines give relatively larger oscillator strengths,
but this is an artifact of the projected wavefunctions.  The 
absorption spectra calculated from the projected wavefunctions
do not satisfy the sum rule, i.e., the area between the abscissa
and the curve does not become constant regardless of the 
excitation wavefunctions.  The similar artifact will be found
in the figures shown afterwards.   believe that the projected
optical absorption spectra are reliable in low energy regions
only.  Therefore, we limit our comparison of the spectra to
the energy region lower than about $2t \sim 4$ eV.

In $\soc$, the edges of the pentagons are the long bonds,
and the bonds between the neighboring hexagons are short
bonds.  The wavefunctions of the fivefold degenerate 
highest-occupied-molecular-orbital (HOMO) have the bonding 
property, and that the threefold degenerate 
lowest-unoccupied-molecular-orbital (LUMO) has the 
antibonding property.  As the carbon number 
increases, hexagons are inserted among pentagons.  The 
wavefunctions near the LUMO of the higher fullerenes still 
have the antibonding properties, thus they tend to have 
large amplitudes along the edges of pentagons which have 
the characters like long bonds of $\soc$.  Recently, the 
bunching of the six energy levels higher than the LUMO  
has been discussed in the extracted higher fullerenes [16].  
The wavefunctions near the LUMO distribute on the pentagons.  
This fact can be understood as the properties characteristic 
to antibonding orbitals.  As the excited electron mainly 
distributes at the pentagonal carbons, the electron-hole 
excitation has large amplitudes at these pentagons.  Thus,
the oscillator strengths of the low energy excitations are
mainly determined by wavefunctions at the pentagonal carbons.
This is the reason why the projected absorptions nearly
accord with the total absorptions in the energy regions
smaller than about 2$t$.

If the projections are performed onto each pentagon, we can
know contributions to optical spectra from the projected carbon
sites.  We would like to look at this feature, for example,
in $D_2$-C$_{76}$.  There are three carbon atoms, which are
not equivalent with respect to symmetries, in this isomer.
These pentagons are indicated by the symbols, A-C, in Fig. 3(a).
The projected absorption spectra are shown by thin curves,
superposed with the total absorption in Figs. 3 (b-d).
The projected absorption is multiplied by the factor 12, 
in order to compare with the total absorption.  We find 
that the projected absorption exhibits small structures 
in the energy region smaller than $2t$.  The structures 
depend on the kind of carbons.  The spectral shapes and 
oscillator strengths are much far from those of the total 
absorption.  It would be difficult to assign experimental 
features of the total absorption with a set of the limited 
number of carbon atoms.  The excitation wavefunctions at 
the twelve pentagons give rise to the shape of the 
absorption spectra totally.

\vspace{1cm}

\noindent
{\large {\bf 2. Nonlinear Optical Response}}\\
In this section, we investigate nonlinear optical properties of 
higher fullerenes.   We focus on the off-resonant third order 
susceptibility in order to estimate the magnitudes of the 
nonlinear optical responses of each isomer.  The Coulomb 
interaction strengths are also changed in a reasonable range, 
because realistic strengths are not well known in higher 
fullerenes.  Based on our results for the optical properties of 
C$_{60}$ and C$_{70}$ [8,9], we can assume $V = U/2$.  
The onsite Coulomb strength is varied within the range 
$0 \leq U \leq 4t$, $t$ being the hopping integral between
nearest neighbor carbon atoms.

In Fig. 4, the relations between the absolute value of the 
off-resonant susceptibility and the energy gap are shown for 
three Coulomb interaction strengths: $U = 0t$, $2t$, and $4t$.  
Here, the energy gap is defined as the optical excitation 
energy of the lowest dipole allowed state, in other words, 
the optical gap.  For each Coulomb interaction, the plots 
(squares, circles, or triangles) cluster in a bunch.  When the 
energy gap becomes larger, the susceptibility tends to decrease.  
However, the correlation between the susceptibility and the 
energy gap is far from that of a smooth function.  The 
correlation is merely a kind of tendency.  Therefore, the 
decrease in the energy gap of higher fullerenes is one origin 
of the larger optical nonlinearities of the systems.  The 
actual magnitudes of nonlinearities would also be influenced 
by the detailed electronic structures of isomers.

In the calculations for C$_{60}$ reported previously, the 
magnitudes of the THG at the energy zero are approximately 
$1 \times 10^{-12}$ esu in the free electron model, and 
approximately $2 \times 10^{-13}$ esu for $U = 4t$ and 
$V=2t$.  In the present calculations for higher 
fullerenes, the magnitudes are a few times larger than those 
of C$_{60}$.  Thus, the author predicts that nonlinear 
optical responses in higher fullerenes are generally larger 
than in C$_{60}$.  In our previous paper [8], we discussed 
the fact that the local field correction factor is of the 
order of 10 for C$_{60}$ solids.  Since the distance between 
the surfaces of neighboring fullerene molecules in C$_{70}$ 
and C$_{76}$ solids is nearly the same as in C$_{60}$ solids, 
we expect that local field enhancement in thin films of higher 
fullerenes is of a magnitude similar to that in C$_{60}$ systems.

\vspace{1cm}

\noindent
TABLE I.  Coulomb interaction dependence of the power $\alpha$
where $|\chi^{(3)}(0)| \sim A \cdot N^\alpha$.

\mbox{}

\begin{tabular}{cc} \hline \hline
$U$  & $\alpha$ \\ \hline
0$t$ & 5.253 \\
2$t$ & 4.133 \\
4$t$ & 3.536 \\ \hline \hline
\end{tabular}

\vspace{1cm}

It is of some interests to look at carbon number dependence
of the magnitude of the optical nonlinearity of the calculated
isomers in higher fullerenes.  Figure 5 shows $|\chi^{(3)}(0)|$
as functions of the carbon number $N$ for three Coulomb interaction
strengths, $U=0t$, $2t$, and $4t$.  The solid lines indicate
the linear fitting in the logarithmic scale: $|\chi^{(3)}(0)|
\sim A \cdot N^\alpha$.  The powers $\alpha$ for the three Coulomb
interaction strengths are summarized in TABLE I.  When $U=0t$, 
the power $\alpha$ is about 5.  As $U$ increases, $\alpha$
decreases.  It is among 4, when $U \sim 2t$ and $4t$.  This
magnitude of the power 4 agrees with the result of the quantum
chemical calculation of higher fullerenes upto C$_{84}$ [17].
Therefore, we have shown important roles of Coulomb interactions
in nonlinear optical response of higher fullerenes.

Experimental measurements of optical nonlinearities in higher
fullerenes whose carbon number is larger than 70 have not been
reported so much, possibly because of the difficulty in obtaining
samples with good quality and the difficult measurements.  However,
the recent report of the degenerate four-wave-mixing measurement
of C$_{90}$ in solutions [18] indicates the larger optical nonlinearity
than that in C$_{60}$.  The magnitude of $\chi^{(3)}$ is about
eight times larger than in C$_{60}$, and is apparently enhanced
from that of the theoretical predictions: $(90/60)^4 = 1.5^4 = 5.063$.
Therefore, further experimental as well as theoretical investigations
of nonlinear optical properties in higher fullerenes should be
fascinating among scientists and technologists of the field of
photophysics.

\vspace{1cm}

\noindent
{\large {\bf References}}

\noindent
$[1]$ J. P. Hare, H. W. Kroto and R. Taylor, Chem. Phys. Lett. 
177 (1991) 394.\\
$[2]$ S. L. Ren, Y. Wang, A. M. Rao, E. McRae, J. M. Holden, 
T. Hager, K. A. Wang, W. T. Lee, H. F. Ni, J. Selegue 
and P. C. Eklund, Appl. Phys. Lett. 59 (1991) 2678.\\
$[3]$ J. S. Meth, H. Vanherzeele and Y. Wang, Chem. Phys. Lett.
197 (1992) 26.\\
$[4]$ Z. H. Kafafi, J. R. Lindle, R. G. S. Pong, F. J. Bartoli,
L. J. Lingg and J. Milliken, Chem. Phys. Lett. 188 (1992) 492.\\
$[5]$ R. Ettl, I. Chao, F. Diederich and R. L. Whetten,
Nature 353 (1991) 149.\\
$[6]$ K. Kikuchi, N. Nakahara, T. Wakabayashi, M. Honda, H. Matsumiya,
T. Moriwaki, S. Suzuki, H. Shiromaru, K. Saito, K. Yamauchi,
I. Ikemoto and Y. Achiba, Chem. Phys. Lett. 188 (1992) 177.\\
$[7]$ K. Harigaya and S. Abe, Jpn. J. Appl. Phys. 31 (1992) L887.\\
$[8]$ K. Harigaya and S. Abe, J. Lumin. 60\&61 (1994) 380.\\
$[9]$ K. Harigaya and S. Abe, Phys. Rev. B 49 (1994) 16746.\\
$[10]$ K. Harigaya, Jpn. J. Appl. Phys. 33 (1994) L786.\\
$[11]$ M. Fujita, Fullerene Science and Technology, 1 (1993) 365.\\
$[12]$ M. Fujita, R. Saito, G. Dresselhaus and M. S. Dresselhaus,
Phys. Rev. B 45 (1992) 13834.\\
$[13]$ M. Yoshida and E. \={O}sawa, Proc. 3rd IUMRS
Int. Conf. Advanced Materials, 1993.\\
$[14]$ M. Yoshida and E. \={O}sawa, 
The Japan Chemistry Program Exchange, Program No. 74.\\
$[15]$ J. Shumway and S. Satpathy, Chem. Phys. Lett. 211 (1993) 595.\\
$[16]$ S. Saito, S. Okada, S. Sawada and N. Hamada,
Phys. Rev. Lett. 75 (1995) 685.\\
$[17]$ M. Fanti, G. Orlandi and F. Zerbetto,
J. Am. Chem. Soc. 117 (1995) 6101.\\
$[18]$ H. Huang, G. Gu, S. Yang, J. Fu, P. Yu, G. K. L. Wong 
and Y. Du, Chem. Phys. Lett. 272 (1997) 427.\\

\pagebreak

Figures should be requested to harigaya@etl.go.jp.

Caption:

Fig. 1.   Phason lines in higher fullerenes.

Fig. 2.   Molecular structures and theoretical optical
spectra of (a) $D_{5d}$-$\rug$ and (b) $D_2$-C$_{76}$.
In the molecules, the black atoms are along the phason
lines, and the hatched atoms are the pentagonal carbons.
In the absorption spectra, the bold line is the total
absorption, and the thin line is the absorption by 
the wavefunctions projected on the twelve pentagons.
The units of the abscissa are taken as arbitrary, and
the energy is scaled by $t$.  The parameters are
$U=4t$, $V=2t$, and $\gamma = 0.06t$.

Fig. 3.  (a) The molecular structure of $D_2$-C$_{76}$.  The 
symbols, A-C, indicate the symmetry nonequivalent pentagons.
The figures, (a), (b), and (c), compare the absorption 
projected on one of the three pentagons with the total 
absorption.  The bold line is the total absorption, and 
the thin line is the projected absorption.  The units of 
the abscissa are taken as arbitrary, and the energy is 
scaled by $t$.  The data of the thin line are multiplied
by the factor 12.  The parameters are $U=4t$, $V=2t$, 
and $\gamma = 0.06t$.

Fig. 4.  The absolute value of the off-resonant susceptibility
$| \chi^{(3)} (0)|$ for C$_{60}$ and seven isomers of
higher fullerenes, plotted against the energy
gap (shown in units of $t$).  The squares, circles, and
triangles represent results for $U = 0t$, $2t$, and $4t$,

Fig. 5.  The absolute value of the off-resonant susceptibility
$| \chi^{(3)} (0)|$ for C$_{60}$ and seven isomers of
higher fullerenes, plotted against the carbon number $N$.  
The squares, circles, and triangles represent results for 
$U = 0t$, $2t$, and $4t$, respectively.  The left and bottom
axes are in the logarithmic scale.  The solid lines are
the results of the linear fitting in the logarithmic scale: 
$|\chi^{(3)}|\sim A \cdot N^\alpha$.

\end{document}